\documentclass[aps,pra,twocolumn,groupedaddress,showpacs]{revtex4}

\usepackage{latexsym,amsmath,amssymb}
\usepackage{graphicx}

\newcommand{\ket}[1]{
   \ensuremath{| #1 \rangle}}

\begin{document}


\title{Photon number resolution using time-multiplexed single-photon detectors}


\author{M. J. Fitch}
\email[]{michael.fitch@jhuapl.edu}
\author{B. C. Jacobs}
\email[]{bryan.jacobs@jhuapl.edu}
\author{T. B. Pittman}
\email[]{todd.pittman@jhuapl.edu}
\author{J. D. Franson}
\email[]{james.franson@jhuapl.edu}

\affiliation{Johns Hopkins University, Applied Physics Laboratory,
Laurel MD 20723-6099}


\begin{abstract}

Photon number-resolving detectors are needed for a variety of
applications including linear-optics quantum computing. Here we
describe  the use of time-multiplexing techniques that allows
ordinary single photon detectors, such as silicon avalanche
photodiodes, to be used as photon number-resolving detectors. The
ability of such a detector to correctly measure the number of
photons for an incident number state is analyzed. The predicted
results for an incident coherent state are found to be in good
agreement with the results of a proof-of-principle experimental
demonstration.

\end{abstract}


\maketitle{}


\section{Introduction \label{sec:intro}}

There has been considerable interest recently in the development
of photon detectors that are capable of resolving the number $n$
of photons present in an incident pulse. Photon number-resolving
detectors of this kind are needed for a linear optics approach to
quantum computing \cite{Knill:2001,Franson:2002}, for example, and
they may have other applications as well \cite{Gilchrist:2003},
such as conditional state preparation \cite{Kok:2001,Lee:2001}.
Here we describe a simple time-multiplexing technique that allows
ordinary single-photon detectors to be used as photon
number-resolving detectors.  The achievable performance of these
devices is comparable to that of cryogenic devices
\cite{Kim:1999,Takeuchi:1999,Cabrera:1998} that are being
developed specifically for number-resolving applications.
Detectors based on an atomic vapor have also been proposed
\cite{Imamoglu:2002,James:2002}.

    Photon number-resolving detectors can be characterized in part by the probability
$P(m|n)$  that $m$ photons will be detected in a pulse that
actually contains $n$ photons.  For linear optics quantum
computing, it is essential that $P(n|n)$  be as close to unity as
possible.  As one might expect, this requires the single-photon
detection efficiency $\eta$ to be as large as possible in order to
maximize the probability of detecting all of the photons present
in a pulse.  Silicon avalanche photodiodes (APDs) operating in the
visible have a relatively large value of $\eta$ and a small dark
count rate, but they only produce a single output pulse regardless
of the number of incident photons.  The time-multiplexing
technique described here avoids that limitation while retaining
the large value of  $\eta$ and other potential advantages
associated with the use of commercially-available APDs.

    The basic idea of the time-multiplexing technique is to divide the incident pulse
into $N$ separate pulses of approximately equal amplitude that are
displaced in time by a time interval $\Delta t$ that is larger
than the detector dead time $\tau$.  The probability that one of
the divided pulses contains two or more photons becomes negligibly
small for $N \gg n$, in which case the fact that an APD can only
produce a single output pulse is no longer a limitation.

\begin{figure}[b]
\includegraphics*[width=3.1in]{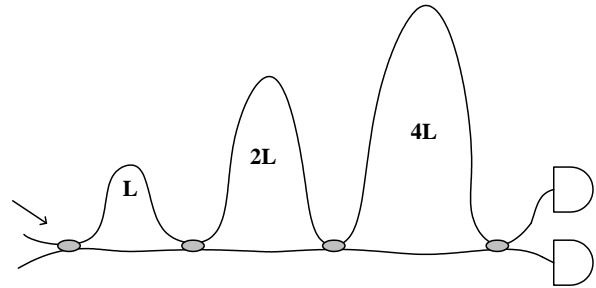}%
\caption{%
Implementation of a time-multiplexed detector (TMD) for photon
number-resolving applications.  An incident pulse is divided into
$N$ weaker pulses separated by a time interval $\Delta t$ using a
series of fiber couplers and optical fiber delay lines.  For
sufficiently large values of $N$ and $\Delta t$, this allows two
ordinary silicon avalanche photodiodes to measure the number of
incident photons with a probability of success that depends on
their detection efficiency. \label{fig:detarrayloops}}
\end{figure}

Our implementation of such a time-multiplexed detector is
illustrated in Figure \ref{fig:detarrayloops}. An incident pulse
propagating in a single-mode optical fiber was divided equally
into two paths using a 50/50 coupler.  The difference $L$ in path
lengths was sufficiently large that the propagation times differed
by more than the detector dead time.  The two pulses were then
recombined at a second 50/50 coupler and split equally into two
more paths differing in length by $2L$, which results in four
pulses all separated in time by $\Delta t$. This process was
repeated once more with a path-length difference of $4L$, after
which a fourth 50/50 coupler split the pulses again and directed
them into one of two silicon APDs.  The net result was the
creation of $N=16$ pulses of approximately equal amplitude. Unlike
earlier approaches involving detector arrays
\cite{Song:1990,Paul:1996,Kok:2001b,Kok:2003}, this technique
requires only two detectors, and it avoids the need for optical
switches and a storage loop
\cite{Banaszek:2003,Rehacek:2003,Haderka:2003}.

The remainder of this paper is organized as follows:  In Section
\ref{sec:theory}, we present a theoretical analysis of the
expected performance of a time-multiplexed detector (TMD) as a
function of the detector efficiency and photon losses.  The
analysis makes clear that high detector efficiency is needed for
any photon number resolving detector, including the TMD and
various cryogenic detectors under development
\cite{Kim:1999,Takeuchi:1999,Cabrera:1998}. Section \ref{sec:expt}
describes an experimental implementation of a TMD and the results
obtained using coherent state inputs.  A summary and conclusions
are presented in Section \ref{sec:sum}.

\section{Theoretical Analysis \label{sec:theory}}

In the most likely application of a time-multiplexed detector, an
incident pulse will contain an unknown number $n$ of photons and
the TMD is intended to measure the value of $n$.  Since the
measurement process destroys any coherence between states with
different values of $n$, it is sufficient to consider an incident
number state (Fock state) with a specific value of $n$ and then
calculate the response of the detector in the form of the
probability distribution $P(m|n)$.  The response to an incident
coherent state will be considered in the next section.

    It is therefore assumed that a number state $\ket{n}=({1}/{\sqrt{n!}}) \,
(\hat{a}^{\dagger})^n \ket{0}$  is incident on the initial 50/50
coupler, where the operator $\hat{a}^{\dagger}$ creates a photon
in the incident mode and $\ket{0}$ is the vacuum state.  The
effects of the first coupler can be described by the operator
transformation
\begin{equation}\label{eq:adagger-trans}
  \hat{a}^{\dagger} \rightarrow (\hat{a}^{\dagger}_{s} +
  \hat{a}^{\dagger}_{l})/\sqrt{2}
\end{equation}
where the operators $\hat{a}^{\dagger}_{s}$ and
$\hat{a}^{\dagger}_{l}$ create photons in the shorter and longer
paths, respectively.  Similar transforms can be applied to
represent the remainder of the 50/50 couplers. After these
transformations, the state vector will involve $n$ photons
distributed over 16 different modes.

It is important to include the effects of photon loss, which will
degrade the performance of the device in a way that is similar to
the effects of limited detector efficiency.  The largest source of
loss is absorption and scattering in the optical fiber. This can
be described by the fraction $f$ of the incident power that is
transmitted through a length $L$ of fiber.  (In the experiments
described in the next section, $f$ has an approximate value of
0.97).  The transmission through the fiber section of length $2L$
is then $f^2$, etc. In order to simplify the analysis, it was
assumed that the short sections of fiber have negligible length
and that losses in the fiber connections are also negligible.

Losses in the optical fiber were included in the analysis by
inserting an additional fiber coupler into each of the longer
loops.  The reflection and transmission coefficients of these
couplers were adjusted to give a transmission probability of $f$
(or the appropriate power of $f$) for each of the sections of
fiber. An additional field mode corresponding to the photons
removed from the loop by these couplers for each of the pulses was
included in the state vector, which increased the total number of
modes from 16 to 23.  The effects of photon loss could then be
taken into account by applying operator transformations analogous
to Eq.\ \ref{eq:adagger-trans}.

After all of the operator transformations had been applied, the
state vector contained a large number of terms corresponding to
all of the ways in which $n$ photons can be distributed over the
23 modes.  As a result, {\sc mathematica} was used to sum the
contribution of each term in the state vector to the probability
distribution $P(m|n)$. In doing so, it was assumed that $P_{0} =
(1- \eta)^q$, where $P_{0}$ is the probability that a detector
will detect no photons if $q$ photons are incident upon it.  (This
ignores any possible correlations between the effects of multiple
photons.) The probability $P_{A}$ that a detector will detect at
least one photon and produce an output pulse is then given by
$P_{A} = 1- P_{0}$. Having determined the values of $P_{0}$ and
$P_{A}$ for each of the detectors, it was straightforward to
calculate the probability of obtaining $m$ detection events and
add that contribution to the probability distribution $P(m|n)$ for
each term in the state vector.

This analysis includes the fact that each of the 16 pulses will
have different probabilities of reaching the detector, since they
all travel through different lengths of optical fiber.  In the
limit of no loss ($f \rightarrow 1$) and perfect detection
efficiency ($\eta \rightarrow 1$), the TMD is mathematically
equivalent to a lossless multiport device, whose output properties
can be obtained analytically \cite{Paul:1996}.  Our numerical
calculations agree with the analytic results in that limit.

\begin{figure}
\includegraphics*[width=3.1in]{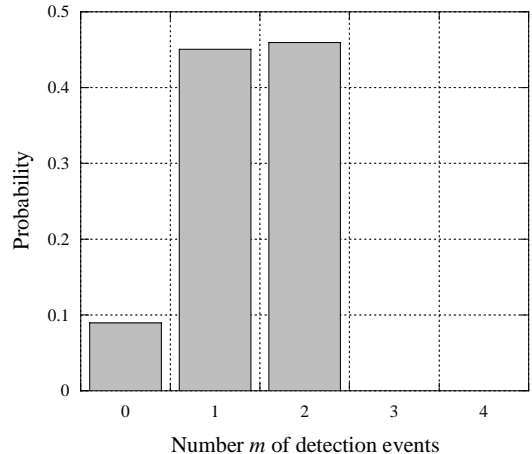}%
\caption{The calculated probability distribution $P(m|n)$ for
observing $m$ detection events given that $n$ photons were
incident on a time-multiplexed detector. These results correspond
to the case of two incident photons ($n=2$) and a detector
efficiency $\eta = 0.7$ with no loss ($f=1$). \label{fig:N2eta7}}
\end{figure}

Figure \ref{fig:N2eta7} shows the results of this analysis for the
case of two incident photons ($n=2$), which is a case of interest
\cite{Knill:2001,Pittman:2001,Pittman:2002,Pittman:2003,Franson:2002}
in linear optics quantum computing.  The detection efficiency was
taken to be $\eta=0.7$, which is typical of commercially-available
silicon APDs, while the losses were assumed to be zero ($f=1$) in
this example. The limited value of $N=16$  reduces the probability
of detecting the correct number of photons to $15/16$, while the
limited detection efficiency further reduces the probability of a
successful measurement to  $\sim$45\%.  It can be seen that a
detection efficiency of 0.7 substantially limits the ability of a
TMD to resolve the number of photons present in a pulse, even for
the simple case of $n=2$, and higher detection efficiencies will
be required in order to improve the performance.

\begin{figure}
\includegraphics*[width=3.1in]{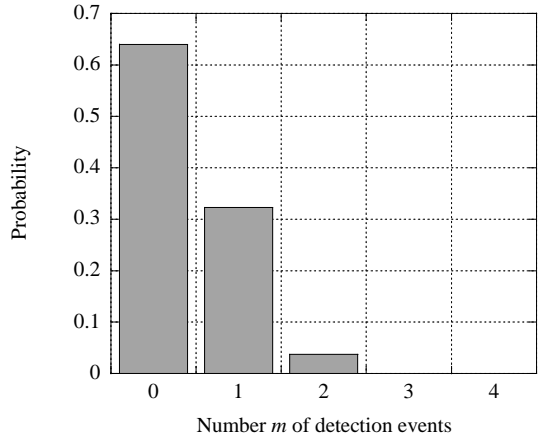}%
\caption{Calculated probability distribution for observing $m$
detection events when two photons are incident and the detection
efficiency is 0.2 with no loss ($f=1$), as in Fig.\
\ref{fig:N2eta7}. \label{fig:2focketa02}}
\end{figure}

Similar results are shown in Figure \ref{fig:2focketa02} for the
case of $\eta=0.2$, which is comparable to the estimated external
efficiency of the superconducting detector in Ref.\
\cite{Cabrera:1998}. Detectors of that kind are mathematically
equivalent to taking the limit of large N, which would increase
the probability of obtaining the correct photon number by 6\%
$(1/16)$ above that shown in Figure \ref{fig:2focketa02}.
Nevertheless, it is apparent that further increases in the
detection efficiency will be required.

\begin{figure}
\includegraphics*[width=3.1in]{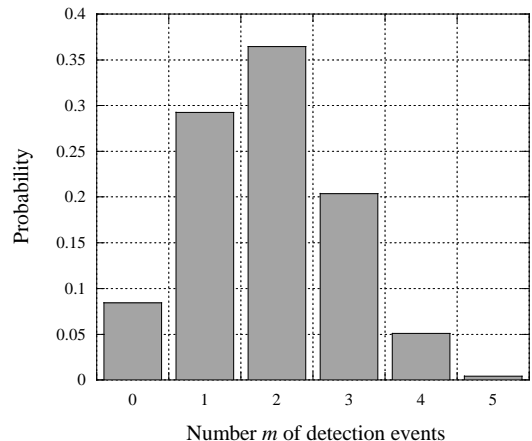}%
\caption{Calculated probability distribution for $m$ detection
events for a Fock state with $n \! = \! 5$. Here the parameter
values were $\eta=0.43$ and $f=0.97$ which correspond to the
experimental conditions of Section \ref{sec:expt} when the
additional losses in the connections and couplers are included.
\label{fig:fock5real}}
\end{figure}

Larger numbers of incident photons give a correspondingly smaller
probability of measuring the correct value of $n$.  As an example,
Figure \ref{fig:fock5real} shows the probability distribution
$P(m|n)$ for the case of $n=5$, $f=0.97$, and $\eta=0.43$.  These
parameters correspond to the experimental apparatus described in
the next section, where the reduced value of $\eta$ includes the
combined averaged effects of input and output coupling loss,
connector and splice losses, and excess losses in the fiber
couplers.

\begin{figure}
\includegraphics*[width=3.1in]{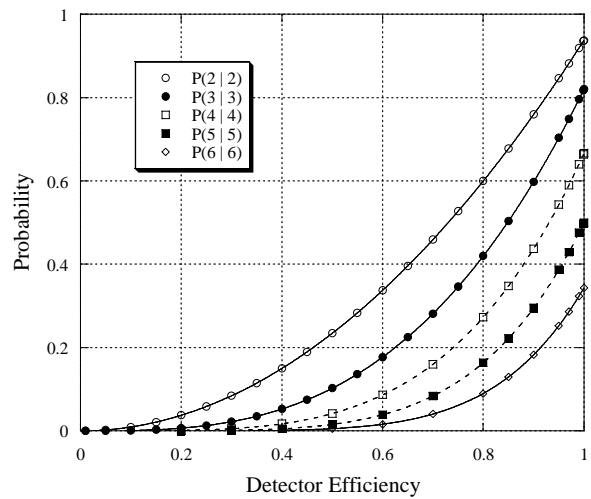}%
\caption{Calculated probability $P(n | n)$ of detecting all $n$
incident photons as a function of the effective detection
efficiency $\eta$. The points were calculated numerically using
the operator transformation technique of Eq.\
\ref{eq:adagger-trans}, while the lines correspond to the analytic
formula of Eq.\ \ref{eq:pmgivenm}. \label{fig:pngivenn}}
\end{figure}

More generally, the probability $P(n|n)$ of measuring the correct
number of photons is plotted in Figure \ref{fig:pngivenn} as a
function of the detector efficiency.  For simplicity, the losses
were assumed to be zero here, in which case $P(n|n)$ has the
analytic result 
\begin{equation}\label{eq:pmgivenm}
  P(n | n) = \frac{16!}{16^{n} (16-n)!} \, \eta^{n} \qquad \text{for} \quad n \leq 16
\end{equation}
as shown in the Appendix. In addition to the effects of limited
detector efficiency, it can be seen that the limited number of
secondary pulses ($N=16$) also has a significant effect as the
value of $n$ increases, even for perfect detector efficiency.  At
telecom wavelengths, the losses in optical fibers are much smaller
and it should be possible to use much larger values of $N$.

\section{Experimental Results \label{sec:expt}}

In order to perform a proof-of-principle demonstration of a
time-multiplexed detector, we implemented the TMD of Figure
\ref{fig:detarrayloops} using single-mode optical fiber and
$2\times 2$ fiber couplers designed to give 50/50 reflection and
transmission at a wavelength of 702~nm. Since a source of number
states is not currently available for large values of $n$, the
experiments were performed using a weak coherent state.  This was
intended to allow a comparison of the theory with experiment and a
demonstration of the basic feasibility of the TMD approach.

The input and output ports of the TMD had FC-style connectors
while all of the other connections were made using a fusion
splicer in order to minimize the losses.  The two detectors were
commercial silicon APD single-photon detectors (Perkin-Elmer model
SPCM-AQR-13) with measured deadtimes of approximately 60~ns.  The
length $L$ of the first long loop was chosen to be 22~m, which
gave a delay between adjacent pulses of $\Delta t = 110$~ns, which
was substantially larger than the deadtime. Short pulses (50~ps
duration) were generated using an externally triggered
fiber-coupled diode laser at 680~nm. A custom-made electronic
circuit was used to count the total number of photon detections in
a specified time window following each trigger.  The triggering
rate was kept sufficiently small that there was no overlap of the
sequence of $N$ pulses from two different triggering events.

The optical fiber used for the delay loops had a 4~$\mu$m core,
which supported single-mode propagation of visible light (610~nm
to 730~nm).  The attenuation at 700~nm was specified by the
manufacturer to be 6~dB/km, which corresponds to a transmission
factor of $f=0.97$. Small but significant losses also occurred in
the input and output couplers, fusion splices, and excess loss in
the $2\times 2$ fiber couplers (typically 3\% each).  The total
transmission of the fiber multiplexing device was measured to be
0.55, averaged over all $N$ pulses.

The photon number distribution of the coherent state at one of the
outputs of the TMD is given by a Poisson distribution:
\begin{equation}\label{eq:poisson-1}
  P(\mu,n)=\frac{\mu^{n} e^{-\mu}}{n!}
\end{equation}
where $\mu$ is the mean photon number. For a perfect detector
($\eta=1$), $P(\mu,0)$ would give the probability $P_{0}$ of
detecting no photons. It can be shown from the linearity of the
system that this result can be generalized to the case of an
imperfect detector by replacing $\mu$ with $\mu^{\prime}=\eta\mu$.
(The limited detection efficiency is equivalent to placing an
attenuator in front of the detector.) Thus $P_{0}=
\exp(-\mu^{\prime})$ and $P_{A}= 1-P_{0}$, where $P_{A}$ is
defined once again as the probability that a detector will detect
at least one photon and produce an output pulse.

For values of $f$ close to 1, the losses for all $N$ pulses can be
taken to be approximately the same without significantly affecting
the distribution of the total number of counts. In that case,
$\mu^{\prime}= \eta l \mu_{o}/16$ for each of the pulses incident
on the detectors, where the transmission factor $l$ accounts for
the average total loss in the system (measured value $l=0.55$),
$\eta$ is the APD detection efficiency (typ.\ 0.7), and $\mu_{o}$
is the mean number of photons incident upon the TMD. The
probability $\mathcal{P}(m)$ of obtaining exactly $m$ detection
events (for $m \leq 16$) is then given by the binomial
distribution
\begin{equation}\label{eq:binomial-click}
  \mathcal{P}(m) = \frac{ 16! }{(16-m)! \, m!} (P_{0})^{16-m}
  (1-P_{0})^{m}.
\end{equation}

\begin{figure}
\includegraphics*[width=3.1in]{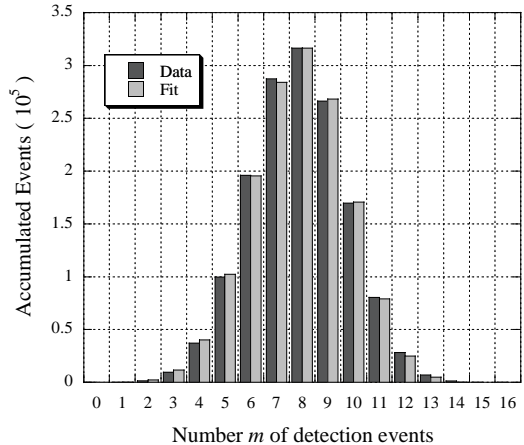}%
\caption{Histogram of the number of events in which $m$ photons
were detected for an incident coherent state pulse with a
relatively large intensity. The light bars correspond to the
theoretical prediction of Eq.\ \ref{eq:binomial-click} based on a
least-squares fit with $\eta l \mu_{o}=13.1$, while the dark bars
correspond the the experimental measurements. \label{fig:nd0fit}}
\end{figure}

A comparison of the theoretical predictions of Eq.\
\ref{eq:binomial-click} with the experimental results from a
relatively intense coherent state are shown in Figure
\ref{fig:nd0fit}.  The data corresponds to a histogram of the
total number of events in which $m$ photons were counted in an
arbitrary data collection interval, which is proportional to the
probability distribution $\mathcal{P}(m)$.  The theoretical
results correspond to a least-squares fit to Eq.\
\ref{eq:binomial-click} with two free parameters, a normalization
constant (which reflects the length of the data collection
interval) and the value of $\mu^{\prime}$, which corresponded to a
best-fit value $\eta l \mu_{o}=13.1$. It can be seen that the data
are in relatively good agreement with the theoretical predictions.
The small discrepancies between the theory and experiment are
probably due to the approximation that all 16 pulses are subject
to the same loss.

\begin{figure}
\includegraphics*[width=3.1in]{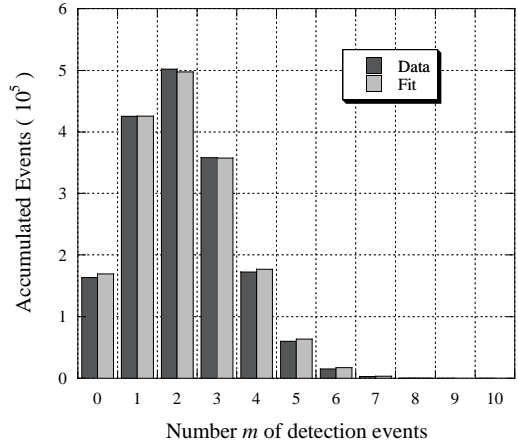}%
\caption{Comparison of the theoretical prediction and experimental
results for an incident coherent state with a smaller intensity
than in Figure \ref{fig:nd0fit}. Here the best fit corresponds to
$\eta l \mu_{o}=2.65$. \label{fig:nd05fit}}
\end{figure}

\begin{figure}
\includegraphics*[width=3.1in]{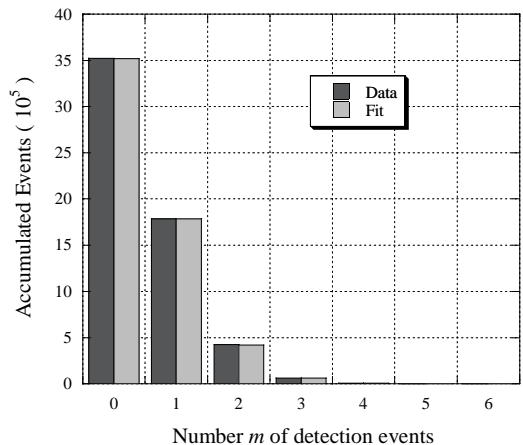}%
\caption{Comparison of the theoretical prediction and experimental
results for an weak coherent state input, as in Figure
\ref{fig:nd0fit}. Here the best fit corresponds to $\eta l
\mu_{o}=0.57$. \label{fig:nd13fit}}
\end{figure}

Similar comparisons of the theoretical predictions and the
experimental results are shown in Figure \ref{fig:nd05fit} for a
less intense coherent state ($\eta l \mu_{o}=2.65$) and in Figure
\ref{fig:nd13fit} for a relatively weak pulse ($\eta l
\mu_{o}=0.57$). It can be seen that a TMD can provide reliable
estimates of the mean number of photons in a coherent-state pulse,
but it can also be seen that the observed peak in the data is
substantially less than the true mean number of photons.

\section{Summary \label{sec:sum}}

In summary, we have described the use of time-multiplexing
techniques to allow ordinary silicon avalanche photodiodes to be
used as photon-number detectors.  Although silicon APDs have high
efficiencies and low dark counts, they only produce a single
output pulse regardless of the number of photons that are
incident.  This difficulty can be avoided by splitting the
incident pulse into a time sequence of $N$ weaker pulses, provided
that $N \gg n$. The use of time multiplexing does lengthen the
effective detector response time, which may also be of importance
depending on the intended application
\cite{Paul:1996,Schleich:1987,Schleich:1988,Dowling:1991}.

    A proof-of-principle experiment was performed using optical fiber loops for the
case of $N=16$ and a coherent state input pulse.  Good agreement
was observed between the theoretical predictions and the
experimental results.

The calculated response of a TMD to an incident number state makes
it clear that high detection efficiencies will be required in
order to truly resolve the number of photons in an incident pulse.
But the same comment applies to any other detector intended for
photon number-resolving applications, such as the cryogenic
detectors that are being developed for that purpose.  The results
of our analysis suggest that the most important requirement for
true number-resolving detectors is a high single-photon detection
efficiency; an intrinsic ability to resolve photon numbers may be
of lesser importance, since that capability can also be achieved
using ordinary single-photon detectors.

    Finally, we would like to note that a similar device has been independently
demonstrated and analyzed by Achilles \emph{et al.}
\cite{Achilles:2003}.

\appendix*
\section{\label{sec:app}}

In this Appendix, we provide a derivation of Eq.\
\ref{eq:pmgivenm}, and generalize it to arbitrary numbers of
detected photons.

Non-classical photon statistics have been observed
\cite{Mattle:1995} at a multiport beam splitter when the path
lengths are matched, and to describe such effects, a quantum
formalism must be used (as in Section \ref{sec:theory} above). In
contrast, the path lengths in the TMD are far from matched, with
relative delays of $>\!100$~ns. As this is two or three orders of
magnitude longer than the coherence length, the possibility of
photon interferences can be neglected, and the relevant detection
probabilities can be calculated  from classical probability
theory.

Let the input mode be evenly distributed to $N$ output modes, so
that a single photon on the input has a probability $1/N$ of
reaching any given output mode. We further assume there are no
losses, corresponding to $f=1$ above, so this model describes a
balanced, lossless $N$-port. At each output mode, there is a
detector with efficiency $\eta$. Suppose that $n$ photons are sent
to the input. Let the probability of $m$ detection events ($0 \!
\leq m \! \leq \! n$) be written $P^{N}_{\eta} (m | n)$.
In the case of zero detections, $P^{N}_{\eta} (0 | n)$ is given
by:
\begin{equation}\label{eq:app-zerodets}
  P^{N}_{\eta} (0 | n) = (1- \eta)^{n}
\end{equation}
which assumes that each photon fails to be detected with an
independent probability of $1-\eta$.  For the case where all
photons are detected, $P^{N}_{\eta} (n | n)$, must include a
probability of $(\eta /N)^n$ for each of the ways in which a
photon can be detected. Since all $n$ photons must have gone to
distinct output modes, there is a combinatoric factor which counts
the number of different ways to distribute $n$ objects among $N$
bins, so that:
\begin{equation}\label{eq:app-ndets}
  P^{N}_{\eta} (n | n) = \left(\frac{\eta}{N}\right)^{n}
  \frac{N!}{(N-n)!} \qquad \text{for} \; n \leq N.
\end{equation}
Eq.\ \ref{eq:app-ndets} reduces to Eq. \ref{eq:pmgivenm} for the
special case in which $N=16$.

For detectors of unit efficiency ($\eta=1$), Paul \emph{et al.}
\cite{Paul:1996} gave a recursion relation for (in our notation)
$P^{N}_{\eta=1} (m | n)$, along with a closed form solution. Their
results can be extended to include non-unit detection efficiency
by using the recursion relation:
\begin{equation}\label{eq:app-recursion}
 \begin{split}
  P^{N}_{\eta} (m | n+1) &= P^{N}_{\eta} (m | n) \left[(1-\eta) +
  \frac{\eta m}{N} \right] \\
  &+ P^{N}_{\eta} (m-1 | n) \left[(N+1-m) \frac{\eta}{N}\right]
 \end{split}
\end{equation}
where the first term on the right hand side describes an
additional photon failing to be detected or else exiting in a
previously occupied output mode. The second term on the right hand
side describes the detection of an additional photon in a
previously unoccupied output mode. In the limit $\eta\rightarrow
1$, the recursion relation Eq.\ \ref{eq:app-recursion} agrees with
that given by Paul \emph{et al.} (see Eq. 9 in Ref.\
\cite{Paul:1996}).

Using the boundary conditions of Equations \ref{eq:app-zerodets}
and \ref{eq:app-ndets}, the recursion relation Eq.\
\ref{eq:app-recursion} can be solved, with the result:
\begin{equation}\label{eq:app-solution}
  P^{N}_{\eta} (m | n) =
  \binom{N}{m} \sum_{j=0}^{m} (-1)^{j} \binom{m}{j}
  \left[(1-\eta) + \frac{(m-j) \eta}{N} \right]^n
\end{equation}
for $m \! \leq \! n \! \leq  \! N$, where the binomial coefficient
is $\binom{N}{m} \equiv N!/[m!(N-m)!]$. The derivation of  Eq.\
\ref{eq:app-solution} is lengthy, but it can be verified that it
satisfies the recursion relation.

Although Eqs.\ \ref{eq:app-zerodets} through \ref{eq:app-solution}
were derived using classical probability theory, they give exactly
the same numerical results as the field operator approach of
Section \ref{sec:theory}.

\begin{acknowledgments}
We acknowledge useful discussions with M. T. Lamar.  This work was
supported by ARO, NSA, ARDA and IR\&D funding.
\end{acknowledgments}

\bibliography{fitch-tmd}

\end{document}